\documentclass[manuscript,screen]{acmart}

\AtBeginDocument{%
  }

\setcopyright{acmlicensed}
\copyrightyear{2025}
\acmYear{2025}
\acmDOI{XXXXXXX.XXXXXXX}

\acmISBN{978-1-4503-XXXX-X/18/06}

\usepackage{graphicx}
\usepackage{textcomp}
\usepackage{xcolor}
\usepackage{colortbl}
\usepackage{adjustbox}
\usepackage{enumitem}
\usepackage{array}
\usepackage{todonotes}
\usepackage{pgfplots}

\begin{document}

\title{Security Modelling for Cyber-Physical Systems: A Systematic Literature Review}

\author{Shaofei Huang}
\email{sf.huang.2023@smu.edu.sg}
\orcid{0009-0002-3954-1843}
\affiliation{%
  \institution{Singapore Management University}
  \city{Singapore}
  \country{Singapore}
}

\author{Christopher M. Poskitt}
\email{cposkitt@smu.edu.sg}
\orcid{0000-0002-9376-2471}
\affiliation{%
  \institution{Singapore Management University}
  \city{Singapore}
  \country{Singapore}
}

\author{Lwin Khin Shar}
\email{lkshar@smu.edu.sg}
\orcid{0000-0001-5130-0407}
\affiliation{%
  \institution{Singapore Management University}
  \city{Singapore}
  \country{Singapore}
}

\renewcommand{\shortauthors}{Huang et al.}

\begin{abstract}
  Cyber-physical systems are at the intersection of digital technology and engineering domains, rendering them high-value targets of sophisticated and well-funded cybersecurity threat actors. Prominent cybersecurity attacks on CPS have brought attention to the vulnerability of these systems and the inherent weaknesses of critical infrastructure reliant on them. Security modelling for CPS is an important mechanism to systematically identify and assess vulnerabilities, threats, and risks throughout system life cycles, and to ultimately ensure system resilience, safety, and reliability. This survey delves into state-of-the-art research on CPS security modelling, encompassing both threat and attack modelling. While these terms are sometimes used interchangeably, they are different concepts. This paper elaborates on the differences between threat and attack modelling, examining their implications for CPS security. We conducted a systematic search that yielded 449 papers, from which 32 were selected and categorised into three clusters: those focused on threat modelling methods, attack modelling methods, and literature reviews. Specifically, we sought to examine what security modelling methods exist today, and how they address real-world cybersecurity threats and CPS-specific attacker capabilities throughout the life cycle of CPS, which typically span longer durations compared to traditional IT systems. This paper also highlights several limitations in existing research, wherein security models adopt simplistic approaches that do not adequately consider the dynamic, multi-layer, multi-path, and multi-agent characteristics of real-world cyber-physical attacks.
\end{abstract}




\maketitle

\section{Introduction}
\label{sec:introduction}

Cyber-physical systems (CPS)~\cite{Griffor2017} are facing increasingly sophisticated cybersecurity threats. Unlike traditional cybersecurity threats, modern cybersecurity threats are volatile, uncertain, complex, and ambiguous (VUCA). These threats are exacerbated by the wide interconnectivity of CPS in smart city infrastructure and the cloud. This interconnectivity presents a myriad of cybersecurity risks from both physical and cyber domains, including unconventional attack paths. In one such example, Kohler et al.~\cite{Kohler2023} demonstrated that the Combined Charging System, a widely used DC rapid charging technology for electric vehicles (EVs), could be exploited wirelessly to interrupt charging sessions for individual vehicles or entire fleets at the same time.

The severity of cyber-attacks on CPS arises from their central role in monitoring, controlling, and optimising critical operations that underpin cities and communities. As these systems sit at the intersection of digital technology and engineering, they have become prime targets for increasingly sophisticated threat actors. Prominent incidents~\cite{Kumar2022} have highlighted not only the vulnerabilities of CPS, but also the inherent weaknesses of critical infrastructure reliant on them. When cybersecurity is not adequately addressed during system design, the consequences of these attacks may be unpredictable and challenging to mitigate, posing considerable risks to public safety, economic stability, and national security.

To address these challenges, CPS owners and operators can take guidance from internationally recognised cybersecurity frameworks and standards. Firstly, the National Institute of Standards and Technology (NIST) Cybersecurity Framework \cite{NationalInstituteOfStandardsAndTechnology2024} provides guidance to identify and manage cybersecurity risks in a systematic way. Secondly, the International Electrotechnical Commission (IEC) 62443 series of standards \cite{IEC62443} include process requirements for the secure development of products used in industrial automation and control systems, offering valuable guidance for securing CPS. These alone are not sufficient to mitigate ever-evolving cybersecurity challenges facing CPS. The dynamic threat landscape requires a tailored approach to CPS cybersecurity, combining continual assessment and adaptation of physical measures, cybersecurity practices, as well as risk mitigation controls.

To situate the discussion in this paper, we define three key terms: ``security modelling'', ``threat modelling'', and ``attack modelling''.
Security modelling refers to the overall process of identifying, assessing, and mitigating risks to protect a system or organisation from a wide range of potential threats, both cyber and physical, throughout its entire life cycle. Within the security modelling context, threat modelling is a proactive approach used primarily in the early stages of a system's life cycle to anticipate potential cybersecurity threats and vulnerabilities, enabling the incorporation of relevant mitigation controls into the system design \cite{Dhillon11,UzunovF14,Xiong2019}. In contrast, attack modelling is a reactive approach that focuses on analysing specific attacker tactics, techniques, and procedures—including both cyber and physical actions—once the system is in operation, to develop targeted defences and response strategies \cite{Al-MohannadiMNA16,Saini2008,Schneier1999,Zenitani23}. To effectively mitigate risks in CPS, integrating threat and attack modelling into a continuous, iterative security modelling process is essential throughout the system’s life cycle \cite{Paudel2017,Yang2023}, allowing systems to adapt to the evolving operational context of CPS environments and the tactics, techniques, and procedures (TTPs) of sophisticated adversaries.

In this paper, we design and conduct a systematic literature review (SLR) to study how existing security modelling, threat modelling, and attack modelling approaches may be integrated in an iterative framework tailored to CPS in a continuous life cycle context. Such a framework is crucial to ensure that CPS security measures are not static but evolve with the changing cybersecurity threat landscape. By bridging this gap, our survey aims to propose a holistic approach to enhance the cyber resilience of CPS.

\textit{Findings}.
In this review, we identified several state-of-the-art security modelling, threat modelling and attack modelling studies relevant to CPS, with the following findings:

\begin{itemize}
    \item Threat modelling is typically conducted in the early stages of CPS development, and may not anticipate new attacker tactics, techniques, and procedures (TTPs) that emerge in later stages of system life cycles.

    \item Security models developed for IT systems in the literature may be difficult to use when modelling CPS cybersecurity threats and attacks. This presents a pertinent challenge to practitioners, given the multi-layer, multi-path or multi-agent characteristics of real-world CPS attacks.

    \item There is limited differentiation in cybersecurity breaches between IT systems and CPS. Unlike IT systems, cybersecurity incidents in CPS may result in complex failure modes, with consequences affecting both cyber and physical domains. Adopting a consequence-driven and cyber-informed approach to CPS security is essential to ensure that both cyber and physical attacks, as well as their consequences, are considered in security modelling.

    \item Unclear definitions and relationships between threat modelling and attack modelling may lead to inconsistent security modelling approaches. We recommend a unified security modelling approach, integrating threat modelling, attack modelling, and security monitoring, to enhance the cyber resilience of CPS and address these challenges.

\end{itemize}

The rest of the paper is structured as follows. Section~\ref{sec:background} provides background on the unique characteristics of CPS security and explains why a tailored approach is essential for securing CPS. Section~\ref{sec:related_work} reviews related work in the field. Our methodology, results, key findings, and limitations are then presented in Sections~\ref{sec:methodology}, \ref{sec:results}, and \ref{sec:discussion}, respectively. In Section~\ref{sec:recommended_approach}, we examine real-world cybersecurity threats through a case study on solar power systems and recommend a CPS security modelling approach to address the gaps identified in both the literature review and the case study. We also discuss the potential applications and associated challenges of this approach across different use cases. Section~\ref{sec:next_steps} consolidates the insights from the literature review and case study, identifies key challenges that remain, and proposes possible directions for future research to address the limitations of current CPS security modelling approaches. Finally, Section~\ref{sec:conclusion} concludes the paper.

\section{Background}
\label{sec:background}

This section provides background on the differences between CPS and conventional IT systems, and the characteristics of adversaries who target CPS in cyber attacks. Together, this explains why a tailored security modelling approach is necessary for CPS.

\subsection{Cyber-Physical Systems}
\label{subsec:cyber_physical_systems}

Different from traditional IT systems, a CPS is a System of Systems (SoS) that integrates computational elements, communication networks, and physical processes to form a unified system. A CPS converges Information Technology~(IT) and Operational Technology (OT) and has different priorities for cybersecurity. OT focuses on safety and reliability, while IT emphasises the confidentiality, integrity, and availability of information. Additionally, CPS networks must accommodate diverse communication modes, ranging from standalone to highly networked systems, potentially utilising legacy protocols such as serial communications, TCP/IP, or object exchange protocols. The heterogeneity and, in certain instances, expansive geographic scopes of CPS networks contribute to the complexity of modelling security for CPS.

Despite their longer system life cycles, CPS are not upgraded or patched frequently as system or software changes risk impacting reliability or safety \cite{KavallieratosKG20,Kriaa2015,SabaliauskaiteM14,Suo2018}. This results in outdated software, exposing CPS to an ever-increasing number of cybersecurity vulnerabilities over time. At the same time, some CPS employ “security-by-obscurity”, using proprietary software and technology that may contain undisclosed security weaknesses vulnerable to exploitation by attackers. Moreover, the lack of “security-by-design” in CPS further reduces the resilience to cybersecurity attacks, as CPS may lack security logs and fail to leverage log data for proactive cybersecurity monitoring and defence. Finally, there is a misconception that merely physically isolating CPS from external networks, known as an "air gap" network security measure, is sufficient to protect CPS from all cybersecurity threats. However, this is a flawed assumption, especially when multi-agent, multi-path CPS attacks use social engineering techniques, physical access, insider threats, or supply‑chain vulnerabilities to breach the "air gap" and compromise CPS even when physically segregated from external networks.

\subsection{Adversarial model}
\label{subsec:adversarial_model}

MITRE ATT\&CK (Adversarial Tactics, Techniques, and Common Knowledge) \cite{Strom2018} serves as a widely adopted knowledge base for characterising the actions and behaviours of cyber adversaries. Specifically tailored for Industrial Control Systems (ICS), i.e.~CPS that automate industrial processes, MITRE ATT\&CK provides a matrix known as "ATT\&CK for ICS" \cite{Alexander2020}. This matrix focuses on the unique tactics, techniques, and procedures (TTPs) employed by adversaries targeting industrial control environments. These TTPs play a crucial role in informing and guiding cybersecurity defence for CPS across the various stages of a cyber-attack life cycle, as outlined in the Lockheed Martin Cyber Kill Chain model \cite{Hutchins2011}.

By analysing past cyber attacks on CPS, Zhu et al.~\cite{Zhu2011} developed a taxonomy that highlighted differences from those observed in IT systems, necessitating a tailored security modelling approach for CPS. CPS attackers employ dynamic, multi-agent, and multi-path attack TTPs to defeat physical access controls and "air gap" measures (isolating CPS from external networks). Additionally, adversaries evolve their TTPs over time, such as using "low and slow" TTPs to evade detection and prolong access to compromised CPS.

\subsection{CPS cyber-intrusion example}
\label{subsec:cyber_intrusion_example}

\begin{figure}[t]
    \centering
    \includegraphics[width=1\textwidth]{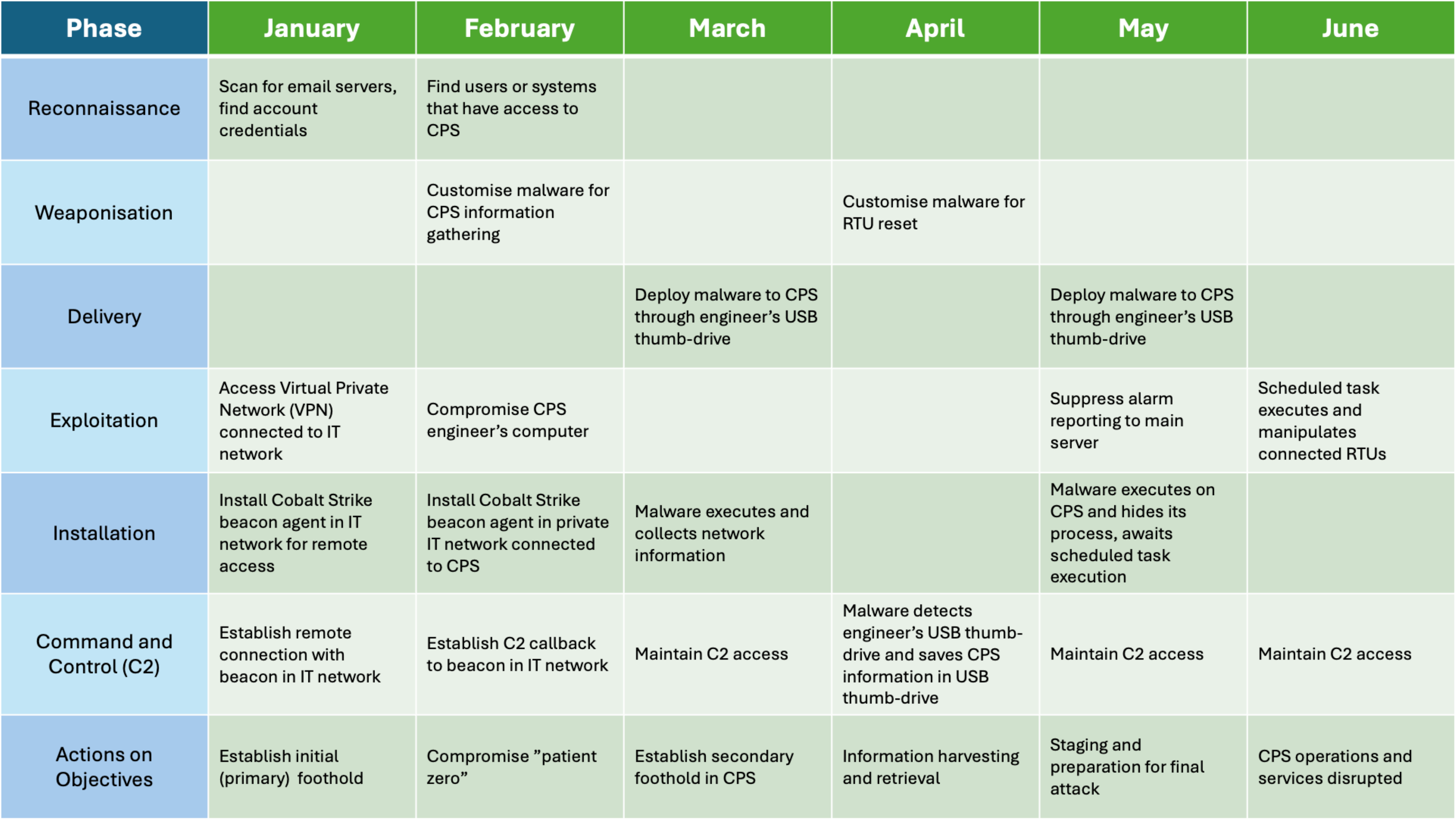}
    \caption{Timeline of a CPS cyber intrusion illustrating the steps in a Cyber Kill Chain \cite{Hutchins2011}}
    \label{fig:intrusion_timeline}
\end{figure}

We present an example of a 6-month long CPS intrusion by a sophisticated attacker group comprising two adversaries, A and B, adapted from the first author's professional experience. In this example, Adversary A compromises the organisation’s IT network to gain remote access to a CPS engineer’s computer and employs social engineering techniques to deliver custom malware to an isolated ("air-gapped") CPS network. Adversary B focuses, on the other hand, on developing custom malware to profile the CPS, and subsequently weaponises it to manipulate connected Remote Terminal Units (RTUs), causing physical damage, for instance via actuators or valves. The Diamond Model of Intrusion Analysis \cite{Caltagirone2013}, as well as a combination of the Diamond Model and the Cyber Kill Chain \cite{Ertaul2018}, are used to describe both the flow (Figure~\ref{fig:intrusion_flow}) and timeline (Figure~\ref{fig:intrusion_timeline}) of this CPS cyber-attack.

This example illustrates the coordinated, multi-stage nature of advanced CPS intrusions, where adversary tactics, targeted malware development, and attack paths across both cyber and physical domains are strategically combined. The tactics, techniques, and procedures employed by sophisticated adversaries are dynamic, evolving in response to both the operational context of CPS environments and the continual emergence of cyber-physical threats and new attack vectors. This highlights the inadequacy of static, one-off security assessments conducted during the conceptualisation or design phases of the CPS life cycle. The example reflects real-world CPS attacks, underscoring the need for iterative and adaptive security modelling approaches that can evolve alongside adversaries and threats, continuously reassess risks, and inform timely defensive measures to safeguard CPS systems.

\begin{figure}[h!]
    \centering
    \begin{tikzpicture}

        \tikzstyle{vertex} = [draw, text width=3.2cm, align=center, minimum height=1.2cm, node distance=1.5cm, font=\scriptsize]
        \tikzstyle{arrow} = [thick,->,>=stealth]

        \node (adversary) [vertex, fill=red!20] {\footnotesize \textbf{Adversary}\\
        - Gains access and establishes persistence within the network};

        \node (infrastructure) [vertex, fill=blue!20, below left of=adversary, xshift=-1.5cm, yshift=-1.6cm] {\footnotesize \textbf{Infrastructure}\\
        - USB thumb drive facilitates malware delivery\\
        - Malware manipulate RTUs, causing CPS disruption};

        \node (capabilities) [vertex, fill=yellow!20, below right of=adversary, xshift=1.5cm, yshift=-1.6cm] {\footnotesize \textbf{Capabilities}\\
        - Malware engineered to inflict physical damage within CPS\\
        - Lateral movement techniques};

        \node (victim) [vertex, fill=green!20, below of=adversary, yshift=-3.9cm] {\footnotesize \textbf{Victim (CPS Engineer)}\\
        - Exploited as intermediary to deliver malware to CPS};

        \draw [arrow] (adversary) -- (infrastructure) node[midway, above, rotate=45] {\scriptsize Uses};
        \draw [arrow] (victim) -- (infrastructure) node[midway, below, rotate=-45] {\scriptsize Delivers};
        \draw [arrow] (capabilities) -- (victim) node[midway, below, rotate=48] {\scriptsize Targets};
        \draw [arrow] (adversary) -- (capabilities) node[midway, above, rotate=-47] {\scriptsize Develops};        \draw [arrow] (capabilities) -- (infrastructure) node[midway, above, rotate=0] {\scriptsize Compromises};

    \end{tikzpicture}
    \caption{CPS cyber intrusion flow representation using the Diamond Model of Intrusion Analysis \cite{Caltagirone2013}}
    \label{fig:intrusion_flow}
\end{figure}

\section{Related Work}
\label{sec:related_work}

There are few existing systematic literature reviews on threat modelling or attack modelling. Humayed et al.'s literature survey \cite{Humayed2017} covered the security and privacy of cyber-physical systems, with a special focus on ICS, smart grids, medical devices, and smart cars. They presented a taxonomy of CPS threats, vulnerabilities, known attacks and existing controls, and importantly, captured how an attack of the physical domain of a CPS can result in unexpected consequences in the cyber domain, and vice versa.

Xiong and Lagerström's SLR \cite{Xiong2019} was another early review on cybersecurity threat modelling, where one CPS threat modelling method was cited: Burmester et al.'s framework \cite{Burmester2012}. Based on the Byzantine paradigm, it enables the modelling of CPS security and facilitates formal analysis and security proof through cryptographic methodologies. Tatam et al. \cite{Tatam2021} reviewed threat modelling approaches for APT-style attacks, but the review did not focus on CPS security. The SLR by Khalid et al. \cite{Khalil2023}, on the other hand, focused on threat modelling of ICS. The authors described the various ICS threat models adopted in the literature, and noted that it was timely to consider a framework that covers all aspects of the various threat modelling methods.

Earlier, the evaluation of threat modelling methods by Shevchenko et al. \cite{Shevchenko2018} reached a similar conclusion: that there is no single method that can cover the full spectrum of CPS threats, and a framework that employs a combination of methods and techniques should be used. Supporting this argument, the hybrid threat model for smart systems proposed by Valenza et al. \cite{Valenza2022}, and the work by Li et al. \cite{Li2021} on security-usability threat modelling for ICS underscored the need to consider not only cyber, but also human factors, owing to the system-of-systems (SoS) nature of CPS.

These CPS threat modelling methods and techniques are well-covered in the literature, with case studies in specific industries and domains, such as the study of smart grid CPS cybersecurity by Nafees et al. \cite{Nafees2023}, the study of cyber-physical energy systems security (CPES) by Zografopoulos et al. \cite{Zografopoulos2021}, the application of the STRIDE threat model in modern vehicle vulnerability assessments by Abuabed et al. \cite{Abuabed2023}, the work on CPS threat modelling for power transformers by Ahn et al. \cite{Ahn2021}, the work on medical CPS threat modelling by Almohri et al. \cite{Almohri2017}, the work by Al Asif et al.~\cite{AlAsif2020} on telesurgery system threat modelling, the work by Jbair et al. \cite{Jbair2022} on threat modelling for industrial cyber-physical systems (ICPS) in the manufacturing industry, the work by Du et al. \cite{Du2024} on attack modelling of power grids with permanent magnet synchronous generator (PMSG)-based wind farms, and the work regarding threat modelling in smart firefighting systems by Zahid et al. \cite{Zahid2023}.

\section{Methodology}
\label{sec:methodology}

\begin{figure}[h]
    \centering
    \includegraphics[width=0.8\textwidth]{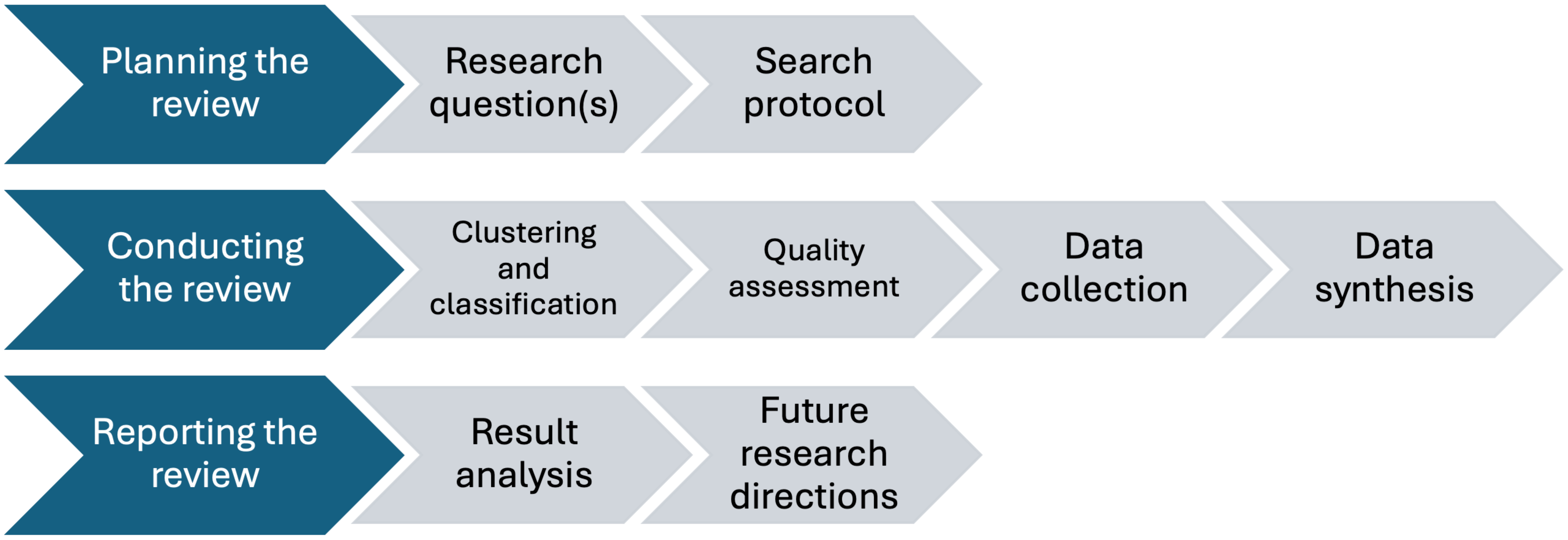}
    \caption{Systematic literature review process}
    \label{fig:literature_review_process}
\end{figure}

This section outlines the methodology used for our systematic literature review (SLR). We followed the guidelines proposed by Kitchenham et al.~\cite{Kitchenham2007}. The process for our SLR is illustrated in Figure~\ref{fig:literature_review_process}.

\subsection{Research questions}

This review aims to answer the following Research Questions (RQs):

\begin{itemize}

\item \textbf{RQ1:} What CPS threat or attack models have been adopted in existing literature?
\\
\textit{The rationale for this research question is to understand the current state-of-the-art research in this area.\\
}
\item \textbf{RQ1.1:} How do the proposed threat and attack models align with cybersecurity in various phases of the CPS life cycle?
\\\textit{The rationale for this research question is to understand how existing CPS threat and attack models address cybersecurity threats not only in the design phase, but when CPS are operational. This is particularly crucial as CPS may operate for many years, often rendering threat models considered during the system design phase obsolete.
}\\
\item \textbf{RQ1.2:} In what manner do the proposed threat and attack models adjust to evolving attacker tactics, techniques, and procedures over time?
\\\textit{The rationale for this research question is to understand how existing CPS threat and attack models are augmented and improved with real-world feedback from cybersecurity intrusions, near-misses, or external threat intelligence information. This is an important bridge between theoretical models and practice, which may result in tangible security improvements for CPS.
}\\
\item \textbf{RQ2:} What research gaps have been identified in the methods used for modelling threats and attacks in CPS?
\\\textit{The rationale for this research question is to identify gaps in current CPS security modelling methodologies and guide future research efforts.
}
\end{itemize}

\subsection{Search protocol}

This sub-section outlines the search protocol used for this review.

We used both US and UK English variations for all search terms. For example, we included both "modelling" (UK English) and "modeling" (US English).

The databases selected for our search were the IEEE Computer Society Digital Library, Science Direct, ACM Digital Library, and Scopus. These databases were chosen because they are recognised as reputable and reliable repositories of academic literature, each with an extensive collection of publications relevant to cybersecurity.

While multidisciplinary approaches that integrate principles from control systems engineering, safety engineering, and human factors are vital for addressing the intricate interplay between cyber and physical components in CPS environments, they are beyond the scope of this paper. A comprehensive review of the full range of strategies required for effective CPS protection would necessitate insights from various specialised domains, each deserving dedicated attention and expertise.

Our primary focus was therefore on search terms related to cybersecurity modelling, security modelling, threat modelling, and attack modelling, in alignment with the subject of our literature review. Criteria A comprised our initial list of search keywords: "security model," "security modelling," "cybersecurity model," "cybersecurity modelling," "threat model," "threat modelling," "attack model," and "attack modelling." Both "model" and "modelling" were included to maximise the coverage of relevant search results.

As we wanted to narrow our search results to papers on security, threat, or attack modelling, we limited the initial search to article titles. Minor adjustments to the search terms were necessary, depending on the Boolean vocabulary supported by each database.

\begin{itemize}
 
    \item At the time of writing, the IEEE Computer Society Digital Library does not support in-line Boolean operators like "OR" and "AND" in search terms. Therefore, we used the Advanced Search menu, selected “in Document Title,” and conducted repeated searches using individual search keywords, such as "security modelling" and "attack model".

    \item On ScienceDirect, we selected Advanced Search and entered (“cybersecurity model” OR “security model” OR “threat model” OR “attack model”) in the Title field. The Boolean “OR” had to be specified, and there was no need to specify the “modeling” keyword as it is implied by “model” in the search term.

    \item On ACM Digital Library, we selected “Advanced Search”, selected “Title” under “Search Within”, then entered: "security model" OR "security modeling" OR "security modelling" OR "cybersecurity model" OR "cybersecurity modeling" OR "cybersecurity modelling" OR "threat model" OR "threat modeling" OR "threat modelling" OR "attack model" OR "attack modeling" OR "attack modelling".

    \item On Scopus, we selected “Article title” under the “Search within” field, then entered "security model" OR "security modeling" OR "security modelling" OR "cybersecurity model" OR "cybersecurity modeling" OR "cybersecurity modelling" OR "threat model" OR "threat modeling" OR "threat modelling" OR "attack model" OR "attack modeling" OR "attack modelling" in the “Search documents” field.

\end{itemize}

We then refined the search results further by specifying additional keywords: ("cyber physical", CPS, SCADA, ICS, APT), to search within article contents. We included SCADA (Supervisory Control and Data Acquisition), ICS (Industrial Control System), and APT (Advanced Persistent Threat) to ensure that the search results were relevant to the adversarial tactics, techniques, and procedures associated with CPS cyber attacks. This list of search terms constitutes Criteria B.

\begin{itemize}
 
    \item On the IEEE Computer Society Digital Library, we used the Advanced Search menu with the initial keywords, selected the “AND” and “in Document Content” options, and entered the additional keywords. Duplicate results arising from the use of varied additional keywords were subsequently identified and removed.

    \item On ScienceDirect, we selected Advanced Search and entered ("cyber physical" OR CPS OR SCADA OR ICS OR APT) in the “Find articles with these terms” field.

    \item On ACM Digital Library, we selected “Edit Search”, added a new search field, selected “Full text” under “Search Within”, then entered: "cyber physical" OR CPS OR SCADA OR ICS OR APT.

    \item On Scopus, we selected “Add search field”, selected “Article title, Abstract, Keywords” under the “Search within” field, and entered: "cyber physical" OR CPS OR SCADA OR ICS OR APT.

\end{itemize}

In the final phase of the search protocol, we removed duplicate papers from the combined search results by comparing paper titles. We also excluded papers that were not related to the research questions, following a review of their titles and abstracts.

\subsection{Categorisation and classification}

We categorised the selected papers into three distinct clusters, based on their primary focus and content:

\begin{itemize}
    \item Cluster 1 (C1): Papers focused on CPS threat modelling method(s)

Papers are placed in this cluster if their primary focus is on threat modelling, defined as the process of identifying potential threats and vulnerabilities early in the system life cycle to inform the design of mitigation strategies. These papers predominantly discuss frameworks, tools, techniques, or methodologies specifically designed for threat modelling.

    \item Cluster 2 (C2): Papers focused on CPS attack modelling method(s)

Papers are included in this cluster if they primarily focus on attack modelling, which involves analysing attacker tactics, techniques, and procedures (TTPs) to inform defence strategies during the system's operational phase. These papers are centred on understanding the behaviour, capabilities, and goals of attackers and often provide detailed descriptions of attacker models, simulation methods, or real-world case studies of cyber attacks.

    \item Cluster 3 (C3): Papers that are systematic literature reviews (SLRs)

Papers are categorised in this cluster if they are systematic literature reviews that provide a comprehensive review of existing threat and/or attack modelling methods. These papers systematically analyse and synthesise findings from a range of studies, offering insights into the state of research, common practices, gaps, and future directions.

\end{itemize}

Papers were categorised based on a thorough analysis of their content, including their primary objectives, methodologies, and the specific aspects of threat or attack modelling they emphasise. Where papers covered both threat and attack modelling, they were classified according to the dominant theme or focus, based on the extent and depth of discussion related to either method.

Additionally, papers were categorised according to the following criteria:

\begin{enumerate}
    \item Year of publication
    \item Document type, i.e.~conference paper, article, review, book chapter, etc.
    \item Number of citations (based on Google Scholar Citations)
    \item Research question(s) paper is relevant to
\end{enumerate}

\subsection{Quality assessment}

In this phase, we reviewed the papers to ensure they were relevant to CPS security, threat or attack modelling, and removed those papers that did not describe these modelling approaches, CPS applications, or CPS case-studies. We also reviewed bibliographies and included additional papers potentially relevant to the review.

\subsection{Data collection}

We studied the papers closely to understand the background, approaches, and results of the various research studies. Importantly, we sought to identify gaps in the research in terms of security modelling in the context of CPS cyber attacks. We documented our notes from the review in a public repository \cite{Huang2025} for future reference.

\subsection{Data analysis}

In the final phase of the SLR methodology, we synthesised our findings and documented our analysis as well as future research directions.

\section{Results}
\label{sec:results}

\begin{figure}[ht]
    \centering
    \includegraphics[width=1\textwidth]{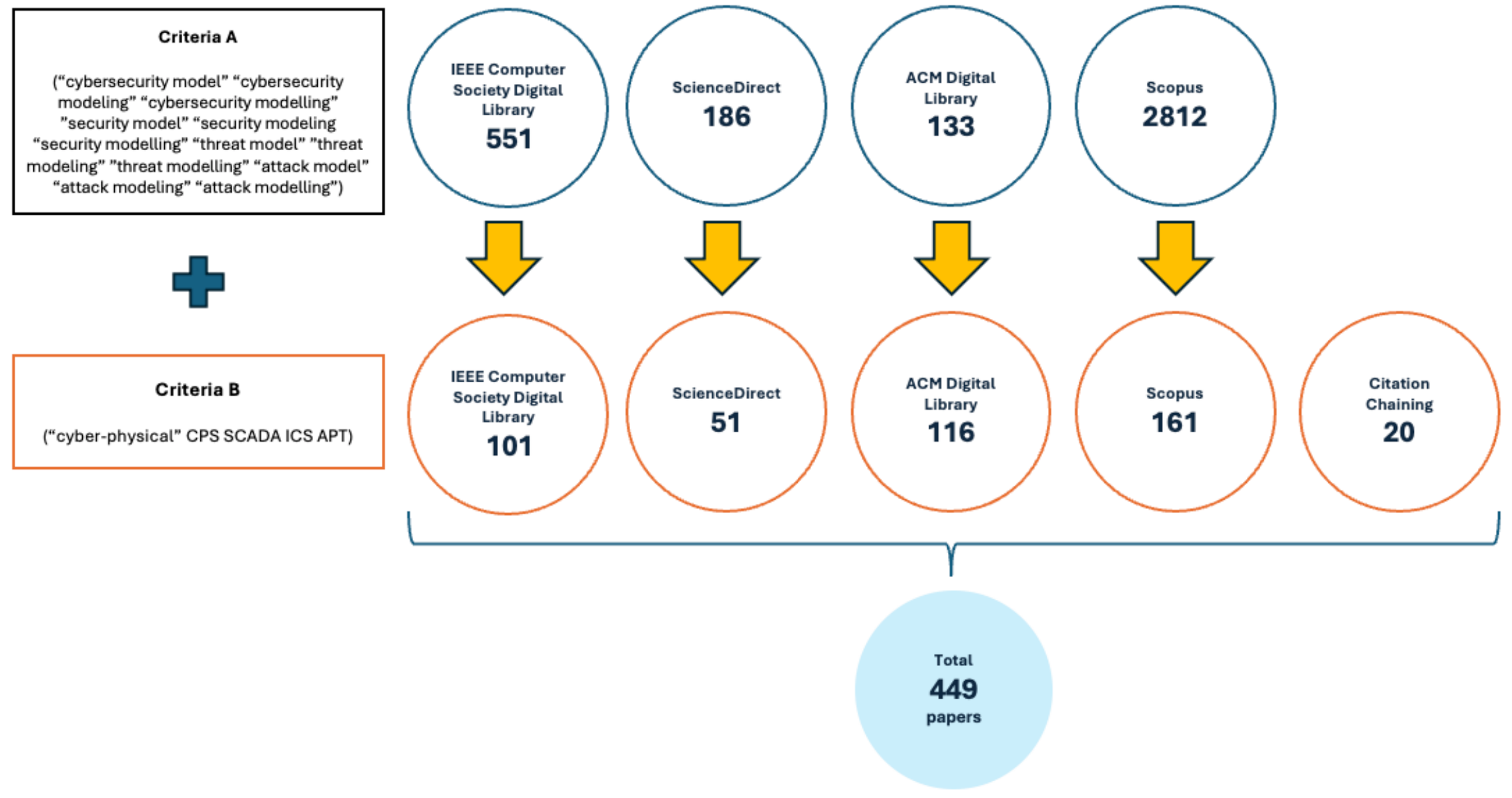}
    \caption{Literature survey search results (April 2025)}
    \label{fig:search_results}
\end{figure}

The search was conducted on the following digital libraries in April 2025: IEEE Computer Society Digital Library, Science Direct, ACM Digital Library, and Scopus. We also adopted alternative search strategies, including citation chaining via Google Scholar and reviewing the bibliographies of screened papers, to identify additional relevant papers. Following the search, we identified a total of 449 papers (Figure~\ref{fig:search_results}). It should be noted that this initial number included some duplicate entries, which were removed in the subsequent step.

To ensure a rigorous quality assessment, we applied a multi-stage filtering process to refine the selection. First, we removed duplicate entries, reducing the count to 350 papers. Next, we conducted an abstract-based screening to eliminate papers that were not relevant to security modelling methods, cybersecurity modelling methods, threat modelling methods, or attack modelling methods, resulting in 125 papers. We then performed a full-text screening to exclude papers that lacked methodological rigour or did not contribute significantly to security modelling research, further reducing the number to 56.

To refine our selection, we categorised the remaining papers based on key criteria, including the security modelling approach (such as threat or attack modelling, formal security modelling, machine learning-based security modelling), the application domain (such as cyber-physical systems, IoT, medical devices), and evaluation methodology (such as conceptual frameworks, case studies, simulation-based approaches). During this process, we removed duplicate papers that presented minor variations of existing models. After this step, we obtained a final set of 32 papers for further analysis. The step-by-step reduction process is summarised in Table~\ref{tab:reduction}.

\begin{table}[h]
    \centering
    \caption{Step-by-step reduction of selected papers}
    \begin{tabular}{lcc}
        \toprule
        \textbf{Step} & \textbf{Remaining Papers} & \textbf{Reduction} \\
        \midrule
        Initial search results  & 449 & - \\
        Duplicate removal & 350 & 99 \\
        Abstract screening (relevance check) & 125 & 225 \\
        Full-text screening (research focus check) & 57 & 68 \\
        Categorisation and overlap removal & \textbf{32} & \textbf{25} \\
        \bottomrule
    \end{tabular}
    \label{tab:reduction}
\end{table}

\begin{table}[]
    \centering
    \caption{CPS cybersecurity papers categorised under threat modelling (C1)}
    \label{tab:categorisation_threat_modelling}
    \small
    \begin{tabular}{|p{3.5cm}|p{9cm}|}
    \hline
    \rowcolor{lightgray}
    Reference & Title \\ \hline
    Ahn et al. (2021) & Security Threat Modeling for Power Transformers in Cyber-Physical Environments \\ \hline
    Badawy et al. (2024) & Legacy ICS Cybersecurity Assessment Using Hybrid Threat Modeling—An Oil and Gas Sector Case Study \\ \hline
    Bhaskar et al. (2024) & A Comprehensive Threat Modelling Analysis for Distributed Energy Resources \\ \hline
    Davis et al. (2024) & Cyber Threat Modeling for Water and Wastewater Systems: Contextualizing STRIDE and DREAD \\ \hline
    Dayarathne et al. (2025) & Mitigating Cyber Risks in Smart Cyber-Physical Power Systems through Deep Learning and Hybrid Security Models \\ \hline
    Fernandez (2016) & Threat modeling in cyber-physical systems \\ \hline
    Fl{\aa} et al. (2023) & A method for threat modelling of industrial control systems \\ \hline
    Khalil et al. (2022) & Threat Modeling of Cyber-Physical Systems-A Case Study of a Microgrid System \\ \hline
    Khan et al. (2017) & STRIDE-based threat modeling for cyber-physical systems \\ \hline
    Kim et al. (2022) & STRIDE‐based threat modeling and DREAD evaluation for the distributed control system \\ \hline
    Martins et al. (2015) & Towards a systematic threat modeling approach for cyber-physical systems \\ \hline
    Mathew et al. (2024) & Hardware-in-Loop (HIL) Testbed Design of Thermal Power Plant for Threat Modeling \\ \hline
    Mekdad et al. (2021) & A threat model method for ICS malware: the TRISIS case \\ \hline
    Noor et al. (2024) & Security and safety in cyber-physical system (CPS): an inclusive threat model \\ \hline
    Saurabh et al. (2024) & TMAP: A Threat Modeling and Attack Path Analysis Framework \\ \hline
    Tang et al. (2024) & ERACAN: Defending Against an Emerging CAN Threat Model \\ \hline
    Valenza et al. (2022) & A hybrid threat model for smart systems \\ \hline
    Zahid et al. (2023) & Threat modeling in smart firefighting systems \\ \hline
    Zalewski et al. (2013) & Threat modeling for security assessment in cyberphysical systems \\ \hline
    \end{tabular}
\end{table}

\begin{table}[]
    \centering
    \caption{CPS cybersecurity papers categorised under attack modelling (C2)}
    \label{tab:categorisation_attack_modelling}
    \small
    \begin{tabular}{|p{3.5cm}|p{9cm}|}
    \hline
    \rowcolor{lightgray}
    Reference & Title \\ \hline
    Du et al. (2024) & Modeling and Assessment of Cyber Attacks  Targeting Converter-Driven Stability of Power Grids  with PMSG-Based Wind Farms \\ \hline
    Iturbe et al. (2024) & A Multi-layer Approach through Threat Modelling and Attack Simulation \\ \hline
    Kumar et al. (2022) & APT attacks on industrial control systems: A tale of three incidents \\ \hline
    Mehmood et al. (2024) & Securing industrial control systems (ICS) through attack modelling and rule-based learning \\ \hline
    Neubert and Vielhauer (2020) & Kill chain attack modelling for hidden channel attack scenarios in industrial control systems \\ \hline
    Paudel et al. (2017) & Attack models for advanced persistent threats in smart grid wide area monitoring \\ \hline
    Yang and Zhang (2023) & From Tactics to Techniques: A Systematic Attack Modeling for Advanced Persistent Threats \\ \hline
    Zhang et al. (2024) & Stochastic important-data-based attack model and defense strategies \\ \hline
    \end{tabular}
\end{table}

\begin{table}[]
    \centering
    \caption{CPS cybersecurity papers categorised under SLRs (C3)}
    \label{tab:categorisation_slrs}
    \small
    \begin{tabular}{|p{3.5cm}|p{9cm}|}
    \hline
    \rowcolor{lightgray}
    Reference & Title \\ \hline
    Ayrour et al. (2018) & Modelling cyber attacks: a survey study \\ \hline
    Khalil et al. (2023) & Threat modeling of industrial control systems: A systematic literature review \\ \hline
    Sa{\ss}nick et al. (2024) & STRIDE-based Methodologies for Threat Modeling of Industrial Control Systems \\ \hline
    Tatam et al. (2021) & A review of threat modelling approaches for APT-style attacks \\ \hline
    Xiong and Lagerström (2019) & Threat modeling – A systematic literature review \\ \hline
    \end{tabular}
\end{table}

Next, the papers were categorised into three clusters based on a thorough reading of the papers: threat modelling (C1) (Table~\ref{tab:categorisation_threat_modelling}), attack modelling (C2) (Table~\ref{tab:categorisation_attack_modelling}), and SLRs (C3) (Table~\ref{tab:categorisation_slrs}).

Out of the selected papers, 19 were categorised under threat modelling (C1) and 8 under attack modelling (C2), indicating that existing literature has focused more on threat modelling methods than attack modelling methods. The earliest selected paper was published in 2013, although no papers were published in 2014. Since then, the number of relevant publications has steadily increased (Figure~\ref{fig:papers_per_year}), reflecting a growing research interest in CPS security modelling. Furthermore, of the 32 selected papers, 18 were journal articles and 14 were conference proceedings. This demonstrates that a greater proportion of research on CPS cybersecurity modelling has been published in journals rather than conferences.

\begin{figure}[t]
    \centering
    \begin{tikzpicture}
        \begin{axis}[
            width=1.0\textwidth,
            height=6cm,
            axis lines=middle,
            ymin=0,
            ytick distance=1,
            xticklabel style = {rotate=45,anchor=east},
            xtick=data,
            grid=major,
            grid style={dashed,gray!30},
            xticklabels from table={pub.dat}{Year},
        ]
            \addplot[orange,thick,mark=square*] table [y=Number,x=Year]{pub.dat};
        \end{axis}
    \end{tikzpicture}
    \caption{Number of CPS cybersecurity papers per year (2013-2024)}
    \label{fig:papers_per_year}
\end{figure}

We used citation metrics on Google Scholar to identify papers that received more than 100 citations. As of April 2025, only 2 of the selected papers received more than 100 citations (Table~\ref{tab:top_cited_papers}). The top cited paper was regarding STRIDE-based threat modelling for cyber-physical systems by Khan et al. \cite{Khan2017}, followed by the SLR on threat modelling by Xiong and Lagerström \cite{Xiong2019}. This observation, together with those from the number of papers published yearly, suggests that this field of research is still relatively new.

\begin{table}[ht]
    \centering
    \caption{Top-cited CPS cybersecurity papers}
    \label{tab:top_cited_papers}
    \begin{tabular}{|l|c|}
    \hline
    \rowcolor{lightgray}
    Reference & Citations \\ \hline
    Khan et al. (2017) & 356 \\ \hline
    Xiong and Lagerström (2019) & 376 \\ \hline
    \end{tabular}
\end{table}

Finally, we performed a thorough reading of the selected papers and analysed which research questions each paper addressed (Table~\ref{tab:paper_relevance}). Three papers \cite{Ayrour2018,Kumar2022,Mekdad2021} addressed all the research questions. The other papers addressed some, but not all research questions. 29 papers were relevant to RQ1, and only 2 of the 19 threat modelling (C1) papers were relevant to RQ1.1 (10.53\%), 1 was relevant to RQ1.2 (5.26\%), and 5 (26.32\%) were relevant to RQ2. On the other hand, 3 of the 8 attack modelling (C2) papers were relevant to RQ1.1 (37.5\%), 4 were relevant to RQ1.2 (50\%) and 2 were relevant to RQ2 (25\%). Lastly, 2 of the 5 SLR (C3) papers were relevant to RQ1 (40\%), 3 were relevant to RQ1.1 (60\%), 1 was relevant to RQ1.2 (20\%), and all 5 were relevant to RQ2.

\begin{table*}[h]
    \centering
    \caption{Paper relevance to research questions}
    \label{tab:paper_relevance}
    \begin{tabular}{|l|c|c|c|c|c|}
    \hline
    \rowcolor{lightgray}
    Reference & Cluster & RQ1 & RQ1.1 & RQ1.2 & RQ2 \\ \hline
    Ahn et al. (2021) & C1 & $\bullet$ &  &  &  \\ \hline
    Bhaskar et al. (2024) & C1 & $\bullet$ &  &  &  \\ \hline
    Davis et al. (2024) & C1 & $\bullet$ &  &  &  \\ \hline
    Dayarathne et al. (2025) & C1 & $\bullet$ &  &  &  \\ \hline
    Fernandez (2016) & C1 & $\bullet$ &  &  &  \\ \hline
    Fl{\aa} et al. (2024) & C1 & $\bullet$ &  &  & $\bullet$ \\ \hline
    Khalil et al. (2022) & C1 & $\bullet$ &  &  & $\bullet$ \\ \hline
    Khan et al. (2017) & C1 & $\bullet$ &  &  &  \\ \hline
    Kim et al. (2022) & C1 & $\bullet$ & $\bullet$ &  &  \\ \hline
    Martins et al. (2015) & C1 & $\bullet$ &  &  &  \\ \hline
    Mathew et al. (2024) & C1 & $\bullet$ &  &  &  \\ \hline
    Mekdad et al. (2021) & C1 & $\bullet$ & $\bullet$ & $\bullet$ & $\bullet$ \\ \hline
    Noor et al. (2024) & C1 & $\bullet$ &  &  &  \\ \hline
    Saurabh et al. (2024) & C1 & $\bullet$ &  &  &  \\ \hline
    Tang et al. (2024) & C1 & $\bullet$ &  &  &  \\ \hline
    Valenza et al. (2022) & C1 & $\bullet$ &  &  & $\bullet$ \\ \hline
    Zahid et al. (2015) & C1 & $\bullet$ &  &  &  \\ \hline
    Zalewski et al. (2013) & C1 & $\bullet$ &  &  & $\bullet$ \\ \hline
    Du et al. (2024) & C2 & $\bullet$ &  &  &  \\ \hline
    Iturbe et al. (2024) & C2 & $\bullet$ &  &  &  \\ \hline
    Kumar et al. (2022) & C2 & $\bullet$ & $\bullet$ & $\bullet$ & $\bullet$ \\ \hline
    Mehmood et al. (2024) & C2 & $\bullet$ &  &  &  \\ \hline
    Neubert \& Vielhauer (2020) & C2 & $\bullet$ &  & $\bullet$ & $\bullet$ \\ \hline
    Paudel et al. (2017) & C2 & $\bullet$ & $\bullet$ & $\bullet$ &  \\ \hline
    Yang and Zhang (2023) & C2 & $\bullet$ & $\bullet$ & $\bullet$ &  \\ \hline
    Zhang et al. (2024) & C2 & $\bullet$ &  &  &  \\ \hline
    Ayrour et al. (2018) & C3 & $\bullet$ & $\bullet$ & $\bullet$ & $\bullet$ \\ \hline
    Khalil et al. (2023) & C3 &  & $\bullet$ &  & $\bullet$ \\ \hline
    Sa{\ss}nick et al. (2024) & C3 & $\bullet$ &  &  & $\bullet$ \\ \hline
    Tatam et al. (2021) & C3 &  & $\bullet$ &  & $\bullet$ \\ \hline
    Xiong and Lagerström (2019) & C3 &  &  &  & $\bullet$ \\ \hline
    \end{tabular}
\end{table*}

A comparison of papers across the clusters (Figure~\ref{fig:comparison_across_clusters}) showed that existing literature focusing on threat modelling (C1) and attack modelling (C2) addressed RQ1 more frequently than the other research questions. Papers focused on attack modelling (C2) appeared to address RQ1.1 and RQ1.2 more than those that focused on threat modelling (C1), highlighting a greater emphasis within attack modelling on cybersecurity considerations throughout the CPS life cycle and on adaptation to evolving attacker tactics.

\begin{figure}
    \centering
    \begin{tikzpicture}
        \begin{axis}[
            ybar,
            xlabel={Cluster},
            ylabel={Relevant papers (\%)},
            xtick=data,
            xticklabels={Threat modelling (C1), Attack modelling (C2), SLR (C3)},
            legend entries={RQ1, RQ1.1, RQ1.2, RQ2},
            legend pos=outer north east,
            enlarge x limits=0.25,
            width=0.8\textwidth,
            height=6cm,
            bar width=0.5cm,
            ymin=0
        ]
        \addplot coordinates {(0,100) (1,100) (2,40)};
        \addplot coordinates {(0,10.53) (1,37.5) (2,60)};
        \addplot coordinates {(0,5.26) (1,50) (2,20)};
        \addplot coordinates {(0,26.32) (1,25) (2,100)};
        \end{axis}
    \end{tikzpicture}
    \caption{Distribution of relevant papers addressing the research questions (RQs) across different clusters. Papers in the attack modelling (C2) cluster demonstrate more coverage of RQ1.1 and RQ1.2 than those in the threat modelling (C1) cluster, highlighting a greater emphasis on cybersecurity considerations throughout the CPS life cycle and adaptation to evolving attacker tactics.}
    \label{fig:comparison_across_clusters}
\end{figure}

\section{Discussion}
\label{sec:discussion}

This section aims to describe the findings and limitations of this review.

\subsection{Findings}
\label{subsec:findings}

The selected papers were categorised into three clusters:  threat modelling methods (C1), attack modelling methods (C2), and SLRs (C3). There were a total of 18 C1 papers, 9 C2 papers, and 5 C3 papers.

\subsubsection{RQ1}

To address the first research question, “What CPS threat or attack models have been adopted in existing literature?”, we examined the threat and attack modelling methods described in the papers, excluding the SLRs. SLRs were excluded because they primarily focus on summarising and synthesising existing research rather than presenting original threat or attack modelling methods. Therefore, they provide an overview of methodologies rather than specific models. The methods we identified include: Microsoft's STRIDE model \cite{Hernan2006}, the Diamond Intrusion Analysis model~\cite{Caltagirone2013}, attack trees \cite{Saini2008,Schneier1999}, attack graphs \cite{Zenitani23}, the Lockheed Martin Cyber Kill Chain \cite{Hutchins2011}, PASTA (Process for Attack Simulation and Threat Analysis) \cite{Ucedavelez2015}, and the MITRE ATT\&CK framework \cite{Strom2018}, as detailed in Table~\ref{tab:threat_attack_models}.

\begin{table*}[ht]
    \centering
    \caption{Threat and attack modelling methods adopted in selected papers}
    \label{tab:threat_attack_models}
    \begin{tabular}{|l|c|p{6cm}|}
        \hline
        \rowcolor{lightgray}
        Reference & Cluster & Threat and Attack Modelling Method(s) \\ \hline
        Ahn et al. (2021) & C1 & STRIDE, CVE-CVSS \\ \hline
        Badawy et al. (2024) & C1 & STRIDE, PASTA, Attack tree\\ \hline
        Bhaskar et al. (2024) & C1 & Cyber Kill Chain, MITRE ATT\&CK \\ \hline
        Davis et al. (2024) & C1 & STRIDE, DREAD \\ \hline
        Dayarathne et al. (2025) & C1 & Deep learning and hybrid security models \\ \hline
        Fernandez (2016) & C1 & Misuse patterns \\ \hline
        Fl{\aa} et al. (2024) & C1 & Model based on IEC 62443 and Data Flow Diagrams (DFDs) \\ \hline
        Khalil et al. (2022) & C1 & STRIDE \\ \hline
        Khan et al. (2017) & C1 & STRIDE \\ \hline
        Kim et al. (2022) & C1 & STRIDE, DREAD \\ \hline
        Martins et al. (2015) & C1 & Generic Modeling Environment (GME) \\ \hline
        Mathew et al. (2024) & C1 & Hardware-in-Loop (HIL) \\ \hline
        Mekdad et al. (2021) & C1 & Diamond Model of Intrusion Analysis \\ \hline
        Noor et al. (2024) & C1 & Fuzzy Inference System (FIS) \\ \hline
        Saurabh et al. (2024) & C1 & STRIDE, MITRE ATT\&CK \\ \hline
        Tang et al. (2024) & C1 & ERACAN \\ \hline
        Valenza et al. (2022) & C1 & TAMELESS \\ \hline
        Zahid et al. (2023) & C1 & MITRE ATT\&CK, NIST \\ \hline
        Zalewski et al. (2013) & C1 & STRIDE \\ \hline
        Du et al. (2024) & C2 & NIST-based security monitoring architecture \\ \hline
        Iturbe et al. (2024) & C2 & Attack graph \\ \hline
        Kumar et al. (2022) & C2 & Attack tree \\ \hline
        Mehmood et al. (2024) & C2 & Attack variant generated through rule-based learning \\ \hline
        Neubert and Vielhauer (2020) & C2 & Cyber Kill Chain \\ \hline
        Paudel et al. (2017) & C2 & Attack tree \\ \hline
        Yang and Zhang (2023) & C2 & Abstract APT attack model \\ \hline
        Zhang et al. (2024) & C2 & Stochastic data-based attack model \\ \hline
        Sa{\ss}nick et al. (2024) & C3 & STRIDE \\ \hline
        \end{tabular}
\end{table*}

\subsubsection{RQ1.1}

On the second research question, “How do the proposed threat and attack models align with cybersecurity in various phases of the CPS life cycle?”, we found that threat modelling was conducted in the early stages of the system development cycle \cite{Khalil2022}, specifically during the system design and validation phases \cite{Khan2017,Kim2022,Martins2015}. Interestingly, unlike threat modelling, none of the C2 papers suggested or mentioned which phase of the system life cycle attack modelling should be performed in. A possible reason is that attack modelling is viewed as a complement to threat modelling methods \cite{Khalil2022,Kumar2022}.

\subsubsection{RQ1.2}

On the third research question, “In what manner do the proposed threat and attack models adjust to evolving attacker tactics, techniques, and procedures over time?”, we found divergent views on how to incorporate attacker tactics, techniques, and procedures (TTPs) into threat modelling frameworks. For example, Khalid et al. \cite{Khalil2022} argued that attack categories in their proposed taxonomy should be abstract and should not include specific details of attack techniques, such as those outlined in MITRE ATT\&CK for Industrial Control Systems \cite{Alexander2020}. In contrast, Fernandez~\cite{Fernandez2016} proposed a threat model using misuse patterns that may include attacker TTPs to represent CPS threats. These differing approaches highlight the challenge of balancing model specificity with adaptability to evolving threats.

Several studies attempted to incorporate attacker TTPs or similar behavioural characteristics into their models. For instance, Mekdad et al. \cite{Mekdad2021} combined the Diamond Model of Intrusion Analysis \cite{Caltagirone2013} and the ICS kill chain \cite{Assante2015} to develop a dynamic ICS threat model. Similarly, attack trees \cite{Schneier1999} and attack graphs \cite{Zenitani23} have been employed to model CPS attacks \cite{Iturbe2024,Kumar2022,Paudel2017}. Neuber et al. \cite{Neubert2020} demonstrated how the Lockheed Martin Cyber Kill Chain \cite{Hutchins2011} could be used to model hidden channel attack scenarios in ICS, while Bhaskar et al. \cite{Bhaskar2024} proposed a hybrid threat modelling approach integrating the Cyber Kill Chain methodology with attacker TTPs from the MITRE ATT\&CK framework. Yang et al. \cite{Yang2023} proposed an abstract APT attack model capable of recognising and correlating attacker TTPs to identify and mitigate complex attacks in ICS. This model emphasises the need to consider a broad range of TTPs to reflect the evolving threat landscape accurately.

However, our analysis revealed that while some models incorporate elements of adaptability to evolving TTPs, there is a significant gap in addressing how these models dynamically adjust to the continually changing tactics of sophisticated adversaries. For example, while models such as those proposed by Mekdad et al. and Yang et al. suggest mechanisms for integrating new TTPs, they do not specify processes for continuously updating or validating these models against emerging threats.

Moreover, while several papers acknowledged the need for models to account for changes over time, they did not explicitly address attacker TTPs as a component of these changes. For instance, Almohri et al. \cite{Almohri2017} noted that the threat model of a medical CPS might change with its corresponding trust model, incorporating time-bound elements such as temporarily trusted individuals or accounts. Similarly, Khalil et al. \cite{Khalil2022} highlighted that systems may evolve over time, necessitating the inclusion of previously excluded information in future threat models to account for system modifications.

Our findings raise the question of whether threat modelling continues to be relevant in CPS security design if carried out only in the early stages of system development, given the typically long development and implementation periods for complex CPS infrastructure. The cyber threats and available technologies to defend against them would likely have changed by the time the CPS becomes operational, often resulting in insecure systems. This issue presents a critical concern potentially affecting national security. CPS are typically used in critical national infrastructure and are therefore prime targets for sophisticated adversaries who, with extended resources, capabilities, and experience, may adopt “low and slow” tactics to infiltrate and maintain persistence in compromised CPS over long periods. Outdated threat models would be unable to adjust and allow the swift detection and eradication of such persistent threats.

This gap suggests a need for more dynamic and resilient security models that can adapt to both the evolving threat landscape and the changing nature of CPS environments.

\subsubsection{RQ2}

For the final research question, “What research gaps have been identified in the methods used for modelling threats and attacks in CPS?”, we made several observations from the literature.\\

\textbf{\textit{Security models primarily focus on IT systems.}}
Firstly, threat and attack models in the literature primarily focus on IT systems, even in the context of CPS security. For example, the STRIDE model developed by Microsoft was used in several papers to identify CPS threats (Table~\ref{tab:threat_attack_models}), notwithstanding that the model is more commonly applied in software development and IT system security. Using STRIDE in CPS threat modelling \cite{Abuabed2023,Badawy2024,Davis2024,Khalil2022,Khan2017,Kim2022,Sassnick2024,Saurabh2024,Zalewski2013} has several limitations that are not well discussed in the literature.

Specifically, the STRIDE approach relies on Data Flow Diagrams (DFDs) to identify system entities and trust boundaries, which has several shortcomings, including inadequate representation of security concepts, data elements, abstraction levels, and deployment information \cite{SionYLBJ20}. Moreover, it is not always possible to construct an accurate DFD for CPS. There are various reasons for this. A CPS is a system-of-systems that includes both cyber and physical components. Availability, accuracy and integrity of data flows between cyber and physical components cannot be assumed. Most importantly, the consequences of CPS compromise extend beyond those described in the STRIDE model, such as physical effects or damage.

This limitation may be addressed by leveraging diverse, multi-disciplinary expertise from physical security and domain specialists to augment the DFD by capturing interdependencies and interactions between cyber and physical components, and not just IT systems. For instance, Fl{\aa} et al. \cite{Flaa2023} proposed a threat modelling method for ICS, incorporating model elements inspired by IEC 62443 \cite{IEC62443} and DFD. This limitation may also be addressed by continually refining the DFD based on real-time data and feedback from system operations and cybersecurity events detected in the CPS environment over time.

Relatedly, the literature refers to CPS use-cases in the context of ICS, power grids \cite{Du2024}, medical CPS (MCPS), SCADA systems, or modern vehicles. We expect that CPS use-cases will increasingly include emerging technologies such as robotics, drones, EV charging infrastructure and smart buildings. Conventional threat modelling approaches, primarily tailored for IT systems, may encounter heightened challenges when addressing cybersecurity threats in these new domains.\\

\textbf{\textit{Attack scenarios are not realistic.}}
Secondly, many threat and attack modelling methods in the literature are based on simplistic cybersecurity attack scenarios, and do not consider the multi-layer, multi-path or multi-agent characteristics of real-world CPS attacks. However, there are notable exceptions. Ahn et al. \cite{Ahn2023} proposed a framework for developing cyber-resilient smart inverters against advanced threat actors. Dayarathne et al. \cite{Dayarathne2025} proposed security models to mitigate cyber risks in smart cyber-physical power systems. Fernandez \cite{Fernandez2016} proposed a patterns-based threat model that considers multiple vulnerabilities exploited in different parts of a distributed system. Similarly, Mathew et al.~\cite{Mathew2024} proposed a threat modelling approach for thermal power plants, while Tang et al. \cite{Tang2024} proposed an approach to defend Controller Area Networks (CANs) in CPS against evolving cybersecurity threats. Zahid et al. \cite{Zahid2023} discussed threat modelling in smart firefighting systems, and Zografopoulos et al. \cite{Zografopoulos2021} analysed Cyber-Physical Energy Systems (CPES) security from a multi-layered attack perspective. In addition, several researchers have proposed techniques to enhance attack models. Mehmood et al. \cite{Mehmood2024} employed rule-based learning, and Zhang et al. \cite{Zhang2024} employed stochastic data methods.\\

\textbf{\textit{Differentiation between IT and CPS cybersecurity breaches.}}
Thirdly, most of the literature did not differentiate cybersecurity breaches in CPS from those in IT systems, even though CPS breaches entail more complex consequences that span both physical and cyber domains. For instance, CPS cybersecurity breaches often result in a gradual degradation of system services before the CPS fails eventually \cite{Zalewski2013}. This is analogous to a system fault in an autonomous vehicle: the vehicle does not come to a stop immediately---which would injure the passengers---but instead gradually slows to a safe stop. Propagation of cyber threats from CPS to IT segments \cite{Noor2024} is also possible. The implication to CPS security is that from a cyber defence perspective, there is an opportunity to remediate intrusions through cyber or physical interventions \cite{Ertaul2018} or to perform dynamic reconfiguration and self-healing tasks \cite{Zografopoulos2021} before the CPS fails completely.\\

\textbf{\textit{Consequence-driven considerations.}}
CPS increasingly face cybersecurity threats that may endanger the safety, reliability and resilience of critical infrastructure. With the convergence of cybersecurity, safety, and reliability in CPS, it would be beneficial to adopt a consequence-driven and cyber-informed approach \cite{Freeman2016} towards threat and attack modelling, and align with system development processes in engineering domains such as the V-model that is commonly used in the automotive industry \cite{Bolz2020}, as well as the failure mode and effect analysis (FMEA) and fault-tree analysis (FTA) techniques used in safety and reliability engineering domains \cite{Kriaa2015}. Doing so would help avoid treating threat or attack modelling as one-off exercises that assume static cybersecurity threats and vulnerabilities, thereby preventing inherent weaknesses from becoming embedded in CPS designs.\\

\textbf{\textit{Distinguishing between CPS threat and attack modelling.}}
Lastly, the literature often used “threat modelling” and “attack modelling” interchangeably, although most references actually discussed threat modelling. Some authors, however, distinguished between the two, promoting clearer definitions of threat and attack modelling \cite{Khalil2023,Xiong2019} or proposing frameworks based on threat models \cite{Burmester2012}. In this paper, we provide explicit definitions for both threat and attack modelling to clarify their distinct roles within cybersecurity (see Section~\ref{sec:introduction}). Based on these definitions and previous studies, we analysed their characteristics and inter-relationships, aligning them with our recommended unified security modelling framework for CPS, as described in Section~\ref{subsec:security_modelling_framework}.

\subsection{Limitations of this review}
\label{subsec:limitations_of_this_review}

Although we aimed to cover as much existing literature on security, threat and attack modelling in CPS as possible, we limited our search protocol to leading scientific databases to ensure the quality of the papers. Furthermore, we limited our search to journals and conference proceedings and consequently, we may have missed relevant papers from other sources such as book chapters.

We also limited the initial scope of the search to keywords in the paper titles. This approach may have excluded papers relevant to security modelling, threat modelling, or attack modelling that did not explicitly mention these terms in their titles. However, to partially mitigate this exclusion, we reviewed the bibliographies of the selected papers and added any other relevant papers that may have been missed initially. Nevertheless, we ensured that we selected only papers that are pertinent to the research questions for eventual review. Based on the review of abstracts of the shortlisted papers from the search, we believe the coverage of the review to be sufficiently comprehensive.

Our research questions were focused on how existing security modelling approaches addressed CPS, and corresponding adversarial tactics, techniques, and procedures. These factors may introduce some bias to the papers selected for review and may have missed papers that are relevant to CPS but did not mention adversarial models specifically. We addressed this by reviewing bibliographies in the selected papers and including additional papers that may be relevant to the review.

\section{Recommended approach for CPS security modelling}
\label{sec:recommended_approach}

In this section, we discuss real-world cyber-physical threats through a case study, recommend a conceptual security modelling framework to address the gaps identified in both the literature review and the case study, and discuss the practical application of this framework, along with the challenges associated with its implementation.

\subsection{Solar power systems: a case study in cyber-physical threats}
\label{subsec:case_study}

To illustrate the challenges and complexities presented by cyber-physical threats, we examine a case study focused on solar power systems. Solar power systems, essential to modern power grids, include components such as solar panels, photovoltaic (PV) inverters, communication dongles, and cloud-based platforms for remote monitoring and control. They also incorporate mobile applications, energy storage batteries, electric vehicle (EV) chargers, and connections to utilities (including substations, transformers, and transmission lines). These interconnected elements create vulnerabilities that leave solar power systems highly exposed to both cyber threats and physical risks.

In March 2025, security researchers discovered vulnerabilities across three vendors of solar PV inverters~\cite{Forescout2025}. The corresponding CVEs of these vulnerabilities are listed below:

\begin{itemize}
    \item CVE-2024-50691, CVE-2024-50684, CVE-2024-50688, CVE-2024-50692, CVE-2024-50690, CVE-2024-50685, CVE-2024-50693, CVE-2024-50689, CVE-2024-50686, CVE-2024-50687, CVE-2024-50694, CVE-2024-50697, CVE-2024-50695, CVE-2024-50698, CVE-2024-50696, CVE-2025-31933, CVE-2025-27939, CVE-2025-27568, CVE-2025-30511, and CVE-2025-24297
\end{itemize}

The researchers proposed an attack scenario similar to that described by Dabrowski et al. \cite{Dabrowski2017}. In Dabrowski et al.'s scenario, an adversary builds (or rents) a botnet of zombie computers and modulates their power consumption in a coordinated way to destabilise the power grid. Similarly, the researchers' scenario involves exploiting and synchronising entire fleets of solar inverters to modulate power demand, thereby disrupting the power grid. Solar inverters, which convert DC electricity generated by solar panels into AC electricity for the grid, are increasingly interconnected and reliant on communication networks for monitoring and control. This interconnection introduces vulnerabilities that can be exploited by attackers.

The criticality of the vulnerabilities in this scenario is exacerbated by the inherent complexity of CPS. Specifically, the consequences of vulnerability exploitation are not confined to cyber aspects, but also include physical aspects due to the interconnection of diverse components within the system and the interactions between cyber and physical components. For example, the communication protocols used to control solar inverters may lack robust security measures, making them susceptible to unauthorised access or manipulation. Additionally, the physical impact of synchronised inverter manipulation, such as sudden fluctuations in power demand, can cascade through the grid, causing instability or even widespread outages.

Several challenges contribute to the complexity of addressing such an attack scenario:

\begin{itemize}
    \item First, the distributed nature of solar power systems means that a large number of devices must be secured, each potentially operating under different configurations and environments.
    \item Second, the dynamic behaviour of CPS, where cyber events can trigger physical consequences and vice versa, complicates the identification and mitigation of risks.
    \item Third, the integration of legacy systems with modern technologies often creates gaps in security, as older components may not have been designed with cybersecurity in mind.
\end{itemize}

These challenges highlight why traditional frameworks may struggle to address such scenarios effectively. Traditional approaches often focus on either the cyber or physical domain in isolation, failing to capture the complex interdependencies between the two. Moreover, they may lack the flexibility to adapt to the evolving threat landscape and the rapid pace of technological advancement. This underscores the need for iterative and dynamic security modelling methodologies that can address the unique challenges of CPS and ensure the resilience of solar power systems throughout their life cycle.

\begin{table}[h!]
    \caption{Comparison of security modelling frameworks}
    \label{tab:comparison_of_security_modelling_frameworks}
    \centering
    \begin{tabular}{|l|p{3.6cm}|p{3.6cm}|p{3.6cm}|}
    \hline
    \textbf{Framework} & \textbf{Description} & \textbf{Strengths} & \textbf{Limitations} \\ \hline
    \textbf{Attack Graph} & Models attack paths as a graph, showing dependencies and sequences of attack steps. & Captures complex attack relationships; useful for automated analysis. & Can be computationally expensive; difficult to scale for very large CPS networks. \\ \hline
    \textbf{Attack Tree} & Represents potential attack paths and objectives in a hierarchical diagram. & Visualises attack scenarios; aids prioritisation of defences based on likelihood and impact. & Can become overly complex for large systems; requires expertise to construct and interpret. \\ \hline
    \textbf{Cyber Kill Chain} & Describes the stages of a cyberattack, from reconnaissance to execution. & Structured view of attack progression; useful for detecting and disrupting attacks. & Focuses on post-attack analysis; less effective for proactive identification of cyber-physical threats. \\ \hline
    \textbf{Diamond Model} & Focuses on the interaction of adversary, infrastructure, victim, and capability to analyse cyber operations. & Provides a multidimensional view of threats, highlighting relationships between key factors. & Primarily analytical; lacks direct application to proactive threat mitigation in design phases. \\ \hline
    \textbf{MITRE ATT\&CK} & A knowledge base of adversary tactics, techniques, and procedures (TTPs). & Comprehensive; maps real-world attack behaviours to defensive strategies. & Primarily IT-centric, does not fully address unique CPS characteristics such as physical threats, real-time constraints, hybrid attacks, or vulnerabilities in legacy systems often found in CPS environments. \\ \hline
    \textbf{STRIDE} & Identifies six categories of threats: Spoofing, Tampering, Repudiation, Information Disclosure, Denial of Service, and Elevation of Privilege. & Helps identify vulnerabilities early in the design phase. & Focuses on theoretical threats and does not provide insights into how they might be exploited, and the attack dynamics in a CPS environment. \\ \hline
    \end{tabular}
\end{table}

As shown in Table \ref{tab:comparison_of_security_modelling_frameworks}, while traditional frameworks contribute to enhancing system security in various ways, none are sufficient in isolation to address the evolving and complex threats faced by CPS. STRIDE helps to identify theoretical vulnerabilities, attack trees visualise potential attack paths, Cyber Kill Chain outlines the stages of an attack for effective response, MITRE ATT\&CK provides real-world insights into adversarial tactics, and the Diamond Model analyses the interplay between adversary, infrastructure, victim, and capability in cyber operations. However, their limitations—including gaps in addressing physical threats, hybrid attack vectors, dynamic changes in CPS environments, and the need for proactive mitigation during design phases highlight the necessity for a more comprehensive and adaptive security modelling approach, such as our recommended framework.

\subsection{A security modelling framework for CPS}
\label{subsec:security_modelling_framework}

Drawing insights from the survey findings and the case study, we recommend an iterative, unified security modelling framework for CPS.

\begin{figure}[ht]
    \centering
    \includegraphics[width=1\textwidth]{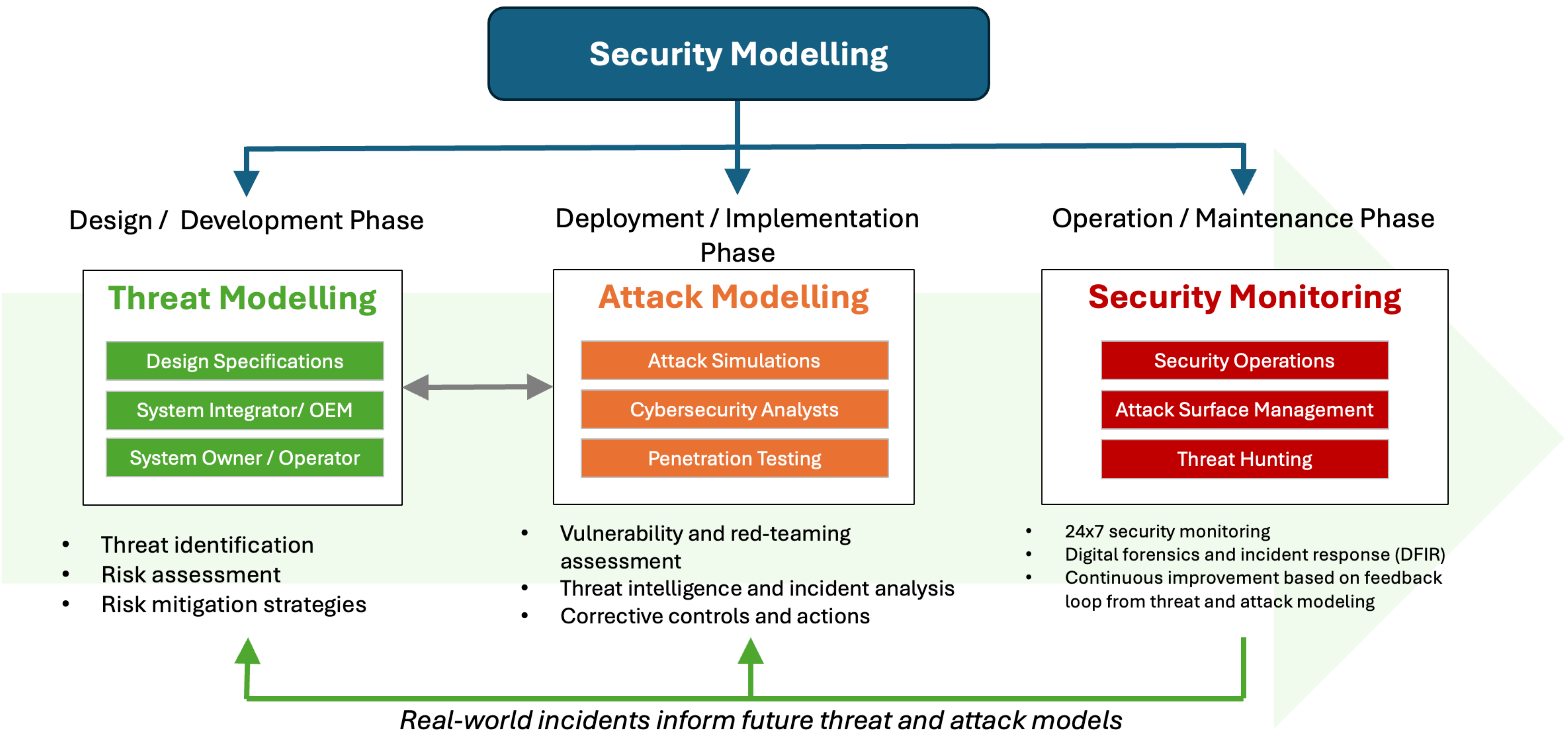}
    \caption{Security modelling framework for CPS that integrates threat modelling, attack modelling and security monitoring throughout the system life cycle}
    \label{fig:security_modelling_framework}
\end{figure}

The framework (Figure~\ref{fig:security_modelling_framework}) begins with threat modelling, a proactive process undertaken during the conceptualisation or design phases to identify potential security risks. This includes analysing attack surfaces, understanding threat actors, and assessing vulnerabilities in hardware, software, and communication protocols. By embedding security-by-design principles early, the system architecture is fortified against anticipated threats. In parallel, attack modelling is employed to validate the effectiveness of implemented security measures against attacker TTPs. Real-world attack scenarios are simulated to assess system resilience, providing valuable insights for refining these measures.

The framework is supported by continuous security monitoring and cyber threat intelligence, ensuring both threat and attack modelling are iteratively updated to reflect newly identified vulnerabilities and attacker TTPs. By implementing this methodology throughout the life cycle of CPS, manufacturers and cybersecurity practitioners can adopt a proactive, risk-informed approach to safeguard these systems effectively.

Our recommended framework is not intended to replace existing threat and attack modelling methods. Instead, it enhances their application by providing a unified and more effective approach for CPS cybersecurity practitioners. The framework enables practitioners to address evolving cybersecurity threats and attacker capabilities comprehensively throughout the CPS life cycle.

\subsection{Applying the framework to solar power systems}

Referring to the case study on solar power systems, the framework could potentially be applied through the following steps:

\begin{enumerate}
    \item \emph{Identify assets:} Begin by ensuring a comprehensive understanding of all relevant assets within the solar power systems, including hardware, software, data, and interfaces (both cyber and physical).

    \item \emph{Construct Data Flow Diagrams (DFDs):} Use tools such as the Microsoft Threat Modelling Tool \cite{MSTMT} to develop and analyse DFDs to map data flows and system interactions. This helps to identify potential cyber-physical threats and attack vectors, including those related to physical system components.

    \item \emph{Model threats (cyber and physical):}
    Use tools such as the Microsoft Threat Modelling Tool to generate cyber-physical threat scenarios based on the DFDs. Additionally, model physical threats—such as unauthorised access to solar power systems or tampering with hardware—to ensure a comprehensive view of the threat landscape.

    \item \emph{Evaluate impact:} Assess the potential consequences of each identified threat, taking into account both operational and safety implications for the solar power systems as well as the grid.

    \item \emph{Model attacks (cyber and physical):} Use attack trees to visualise attacker objectives and possible attack paths, encompassing both cyber and physical vectors.

    \item \emph{Evaluate attack feasibility:} Analyse and quantify the feasibility of each attack using the Common Vulnerability Scoring System (CVSS), considering factors such as exploitability and impact to the solar power systems and the grid.

    \item \emph{Evaluate risks:} Assess the overall cybersecurity risk using a 5×5 risk matrix, combining both the likelihood and impact of the identified threats.

    \item \emph{Review, update, and iterate:} Regularly review the threat model and risk assessment in light of new threat intelligence, insights, or changes to the solar power system environment. Referring to the case study, the risk assessment may be reviewed in light of the newly-discovered vulnerabilities and the corresponding CVEs across the solar PV inverters. This iterative process supports the refinement of mitigation strategies and the strengthening of security controls against emerging threats.

\end{enumerate}

\subsection{Applications and challenges of the framework}

CPS security practitioners can potentially apply the framework in various use cases in the security-by-design, governance, risk management, and compliance domains. For the latter, it can aid practitioners in conducting detailed risk assessments, such as those required under the IEC 62443 series of standards \cite{IEC62443}, which address cybersecurity in industrial automation and control systems (IACS), or the NIST Cybersecurity Framework (CSF) \cite{NationalInstituteOfStandardsAndTechnology2024}.

In the following sections, we outline anticipated opportunities and challenges associated with applying the framework in real-world scenarios. These observations are prospective and based on initial analyses. A comprehensive evaluation of the framework's implementation and effectiveness is beyond the scope of this paper and is identified as an area for future work.\\

\subsubsection{Opportunities and strategies for applying the framework}
\label{subsubsec:opportunities_for_applying_framework}

\paragraph{Risk management.}
By adopting and integrating this framework into organisational processes, organisations may be able to continuously address both operational technology (OT) security requirements and broader organisational risk management objectives for continual monitoring and oversight. For instance, the dynamic security modelling methodology within the recommended framework aligns with and supports organisations in bridging and meeting the requirements of both IEC 62443 and NIST CSF throughout CPS life cycles. While IEC 62443 provides detailed technical and procedural requirements for securing IACS (a subset of CPS), NIST CSF offers a high-level framework for managing cybersecurity risks across diverse organisational contexts that extend beyond IACS or CPS. In this regard, NIST CSF serves as a common organisational structure that integrates multiple cybersecurity approaches by assembling standards, guidelines, and practices, including those specified by IEC 62443.

\paragraph{Governance and compliance.}
The practical benefits of adopting this framework include streamlined compliance with regulatory requirements, reduced redundancy in cybersecurity efforts, and enhanced resilience against evolving threats. By harmonising the requirements of IEC 62443 and NIST CSF, the unified methodology simplifies governance by the organisation's cybersecurity department and helps facilitate compliance with regulatory requirements, enabling organisations to demonstrate adherence to both standards through a single, cohesive approach.

\paragraph{CPS life cycle management.}
The IEC 62443-2-1 standard defines the requirements for establishing a security programme for IACS asset owners, including comprehensive risk assessments and life cycle management. A unified security model enables organisations to systematically implement these requirements by providing a structured approach to identifying and mitigating risks during the system conceptualisation and design phase, as well as later phases throughout the system life cycle. At the same time, it aligns with the NIST CSF’s 'Identify' function, ensuring that risk management processes are consistently integrated into the organisation's overall cybersecurity strategy.

\paragraph{Cross-domain collaboration.}
Additionally, the framework enables CPS practitioners to incorporate threat modelling techniques that integrate safety and security perspectives \cite{Hollerer2021} into their broader cybersecurity risk management framework. This approach could be effective in engaging diverse domain specialists within CPS and in facilitating cross-domain collaboration, fostering a shared understanding of plausible cyber-physical attacks and their impacts on safety and security.

\paragraph{Simulation exercises and training.}
Finally, the framework could be practically applied in cybersecurity incident simulations or wargaming exercises, as well as in security training to dynamically assess the consequences of cyber-physical attacks on CPS and their effects on system resilience and safety. This application could prove invaluable in incorporating threat intelligence and contextual information in a simulation exercise or training setting into the model, thereby supporting decision-making and aligning strategies across affected business units within the organisation.

\subsubsection{Challenges in framework implementation}

\paragraph{Developing dynamic security models.} Dynamic security models rely on accurate baselines, such as up-to-date asset inventories, architecture and network diagrams, and physical layout maps, to construct essential components like Data Flow Diagrams (DFDs). This poses significant challenges in CPS environments, particularly for larger and geographically distributed systems, where asset inventories may be outdated or incomplete. In legacy CPS, this information is often missing altogether, making it difficult to build effective security models. Moreover, the lack of a centralised Configuration Management Database (CMDB) equivalent in many CPS environments exacerbates the problem.

\paragraph{Integrating threat intelligence.} Effective implementation of the framework requires the integration of real-time threat intelligence from sources such as cyber threat intelligence (CTI) feeds, security advisories, and incident reports. This necessitates interoperable data formats and platforms capable of ingesting, transforming, and sharing data for threat and attack modelling. However, correlating external threat intelligence with CPS system logs and data can be complex, especially when the formats are incompatible or require extensive validation to ensure accuracy.

\paragraph{Mitigating cyber-physical vulnerabilities.} In CPS environments, addressing cyber-physical  vulnerabilities is often constrained by operational requirements. Security patches cannot always be applied immediately due to concerns over system availability or disruptions to critical services. Additionally, mitigating vulnerabilities in one system may inadvertently affect the operation of interconnected systems, creating further challenges. CPS software or configuration changes may also lead to system instability or trigger unintended consequences, such as fail-secure mechanisms activating under unexpected conditions. Legacy CPS equipment, which is often end-of-life or end-of-support, lacks the necessary security patches to address newly discovered vulnerabilities, leaving organisations reliant on compensatory controls that may be insufficient in the face of evolving threats.

\paragraph{Resource requirements for legacy CPS environments.}
Implementing the framework in legacy CPS environments requires substantial resources, primarily due to the inherent limitations of outdated systems and the lack of modern infrastructure. Firstly, organisations may need to invest resources to create or update CPS asset inventories, architecture diagrams, and physical layout maps---essential for security modelling. Secondly, resources may be required to integrate security monitoring tools, data interfaces, and threat intelligence platforms with legacy CPS systems, many of which may lack native support for such technologies. This may involve acquiring custom solutions or hardware upgrades to facilitate compatibility. Furthermore, human resources could be a key obstacle to implementation, as specialised expertise in both cybersecurity and CPS operations is needed to implement the framework effectively. Training staff on these updated processes and tools also could demand time and financial investment.

\paragraph{Adapting to diverse CPS environments.} CPS systems are highly diverse, encompassing different operational contexts, technologies, and security architectures. For example, a smart city environment integrates systems such as traffic management, energy distribution, and water supply, all of which operate under distinct technologies and security requirements. This diversity complicates the standardisation of security measures and the implementation of a unified framework. Organisations may tailor their approach to address specific risks and constraints within each environment, which may be resource-intensive and require specialised expertise.

\paragraph{Balancing security and operational requirements.}
Security measures should strike a balance between protecting systems and maintaining their functionality. In CPS environments, particularly those supporting critical infrastructure, overly restrictive controls may hinder operations or affect service delivery. Developing a framework that effectively balances these requirements without compromising security poses a significant challenge.

In conclusion, while our recommended framework could offer potential benefits to CPS practitioners, its implementation requires careful planning and consideration to address challenges such as the development of dynamic security models, integration of threat intelligence, and resource requirements. Overcoming these challenges may enhance the security and resilience of CPS, equipping organisations to better address evolving cyber-physical threats.

\section{Next steps}
\label{sec:next_steps}

In this literature review, we identified several state-of-the-art threat modelling and attack modelling studies relevant to CPS, with the following findings:

\begin{itemize}

    \item Threat modelling is typically conducted in the early stages of system development, and is unlikely to anticipate new attacker tactics, techniques, and procedures (TTPs) in later stages of system life cycles.

    \item Security models in the literature that focus on IT systems could be challenging to use when modelling CPS cybersecurity threats and attacks. This is a pertinent issue for practitioners, given the multi-layer, multi-path, or multi-agent characteristics of real-world cybersecurity attacks in CPS.

    \item There is limited differentiation between cybersecurity breaches in IT systems versus those in CPS. Unlike IT systems, cybersecurity incidents in CPS may result in complex failure modes, as well as consequences in both cyber and physical domains. Adopting a consequence-driven and cyber-informed approach to CPS security is vital for ensuring that cyber and physical attacks and consequences are considered in security modelling.

    \item Unclear definitions and relationships between threat modelling and attack modelling may lead to inconsistent security modelling approaches. A unified security modelling approach that integrates threat modelling, attack modelling, and security monitoring to enhance the cyber resilience of CPS, as recommended in Section~\ref{subsec:security_modelling_framework}, is intended to address this.

\end{itemize}

These findings reveal gaps and limitations in current CPS security modelling approaches. Addressing these gaps requires overcoming several challenges that hinder the development and implementation of effective CPS security modelling frameworks, as detailed below.

\paragraph{Cyber threat intelligence integration}

Developing a dynamic security model presents various challenges, including the integration of real-time cyber threat intelligence into the system's architecture. This demands substantial computational resources for data ingestion, processing, and correlation across diverse CPS components. Furthermore, resource-constrained CPS devices may struggle to support the continuous monitoring and computational overhead required for anomaly detection algorithms.

Incorporating these aspects also requires robust threat intelligence sharing frameworks that address regulatory, ethical, and operational constraints, particularly in multi-stakeholder CPS ecosystems. Potential approaches for integrating threat intelligence into security models include the use of blockchains for secure and immutable sharing of cybersecurity threat intelligence (CTI) \cite{Hajizadeh2020} and federated learning techniques to collaboratively train anomaly detection models while preserving data privacy and ownership \cite{Cui2021}.

\paragraph{Addressing CPS-specific concerns}

Traditional IT security metrics, such as mean time to detection (MTTD) and mean time to resolution (MTTR), often fail to account for the cascading effects of cyber-physical attacks in CPS environments. These metrics do not adequately address CPS-specific concerns, including physical impact, safety implications, and gradual system degradation. To overcome these limitations, security models could incorporate various metrics \cite{He2024}, such as cascading fault graphs (CFG) \cite{Wei2018} to quantify the propagation of failures across both physical and cyber layers of CPS. In addition, safety-critical metrics, such as the probability of breaching physical failure thresholds or safety limits during an attack, could offer valuable insights for addressing CPS-specific vulnerabilities effectively.

Another technical consideration involves scaling these dynamic security models to match the heterogeneity and complexity of CPS environments such as smart grids. Future work could explore the trade-offs between centralised and decentralised security mechanisms in order to balance efficiency, scalability, and latency requirements.

\paragraph{Validation and implementation challenges}

A significant challenge in this area is the validation of anomaly detection approaches, given the difficulty of obtaining high-quality datasets that accurately represent real-world CPS attack scenarios. Many CPS environments are proprietary, and OEMs (Original Equipment Manufacturers) are often reluctant to share data due to intellectual property concerns and regulatory constraints. This lack of access to real-world attack data hinders the development and benchmarking of effective security models. These challenges are further exacerbated by the reliance on legacy technologies that were not originally designed with security in mind. Together, these factors highlight the pressing need to validate and implement modern cybersecurity solutions in CPS environments.\\

To address these identified findings, limitations, gaps, and challenges, we propose several promising directions for future work, as outlined below.

\paragraph{Validating the implementation and effectiveness of our recommended framework}

The various opportunities and strategies for applying our recommended CPS security modelling framework, as outlined in Section~\ref{subsubsec:opportunities_for_applying_framework}, warrant further exploration and evaluation. These include practical applications such as risk management, governance and compliance, CPS life cycle management, cross-domain collaboration, as well as simulation exercises and training.

\paragraph{Modelling CPS intrusions}

To better model CPS cyber intrusions, Petri net based approaches \cite{Chen2011,Dahl2006} may integrate both cyber and physical actions. Petri nets are a graphical and mathematical tool for studying concurrent and distributed systems and have been widely applied in many different areas of computer science and other disciplines \cite{Murata1989}. In the context of CPS intrusion analysis, Petri nets are invaluable for modelling the interactions between cyber and physical components over time, thereby overcoming limitations of existing security modelling approaches when considering dynamic, multi-agent cyber-physical threats.

An alternative approach to modelling CPS intrusions is through Bayesian Belief Networks (BBNs), which provide a probabilistic graphical model for reasoning under uncertainty \cite{Zebrowski2022}. Unlike Petri nets, which primarily model system behaviour in a deterministic or event-driven manner, BBNs incorporate conditional probabilities to represent uncertainty, dependencies, and risk assessment in cyber-physical attacks. BBNs are particularly useful in scenarios where attack likelihood, system vulnerabilities, and detection probabilities are uncertain and need to be estimated based on historical data, expert knowledge, or real-time observations. Additionally, BBNs enable inference-based decision-making, allowing security analysts to evaluate the probability of an ongoing attack given observed system behaviours.

While both Petri nets and BBNs offer valuable insights into CPS intrusion modelling, they differ in their strengths and limitations. Petri nets excel at modelling sequential and concurrent attack processes, making them effective for tracking attack propagation over time. However, they struggle with uncertainty representation. In contrast, BBNs provide a more flexible probabilistic reasoning framework but lack a natural way to model event sequences and real-time attack propagation, which requires additional mechanisms for temporal reasoning.

\paragraph{Exploring self-healing techniques}

Our future work may involve researching self-healing techniques in CPS, to bolster existing cybersecurity defences and strengthen system resilience, whether through cyber or physical actions, against sophisticated adversaries and threats. Self-healing mechanisms could include adaptive response strategies, autonomous system recovery, and AI-driven security orchestration. However, implementing self-healing capabilities within CPS is particularly challenging due to the constraints imposed by legacy infrastructure, safety-critical operational requirements, and regulatory compliance. Many CPS systems operate in environments where downtime is not an option, making the deployment of dynamic security measures more complex. Developing practical self-healing capabilities will further require close collaboration with OEMs, policymakers, and industry stakeholders to align safety-critical requirements, regulatory compliance, and operational feasibility.

\section{Conclusion}
\label{sec:conclusion}

In conclusion, growing interest in and adoption of cyber-physical systems underscores the pressing need for continued innovation in CPS cybersecurity research. As cyber-physical interactions become increasingly complex, developing robust security frameworks is critical for protecting critical infrastructure and reinforcing the resilience of interconnected systems.

Through our systematic literature review, we identified several key challenges and limitations in existing CPS security modelling approaches. First, threat modelling is typically confined to the early stages of system development and may fail to anticipate evolving attacker tactics, techniques, and procedures (TTPs) throughout the CPS life cycle. Second, existing security models -- largely designed for IT systems -- often do not adequately capture the multi-layered, multi-path, and multi-agent nature of CPS attacks, posing practical difficulties for real-world applications. Third, there is limited recognition of the consequences of cyber-physical security incidents in CPS, which may lead to complex failure modes across both cyber and physical domains. Lastly, ambiguities in the definitions and relationships between threat and attack modelling contribute to inconsistencies in current practices.

In response to these challenges, as well as insights gained from a case study focused on solar power systems, we recommend a unified CPS security modelling framework that integrates threat modelling, attack modelling, and security monitoring. This approach aims to enhance the cyber resilience of CPS by providing a more comprehensive and adaptable foundation for managing risks throughout the system life cycle.

\bibliographystyle{ACM-Reference-Format}
\bibliography{references}

\end{document}